\documentclass[aps,twocolumn,preprintnumbers,showpacs,showkeys]{revtex4}
\usepackage{epsfig}
\usepackage{amssymb,amsmath,amsfonts,amsthm,graphicx,psfrag}
\graphicspath{{./Figures/}}                 
\newcommand{\beq}{\begin{equation}}
\newcommand{\eeq}{\end{equation}}
\newcommand{\bea}{\begin{eqnarray}}
\newcommand{\eea}{\end{eqnarray}}
\begin{document}

\preprint{ITEP-LAT/2012-09,~~HU-EP-12/26}

\title{Cooling study of Dirac sheets 
       in $SU(3)$ lattice gauge theory below $T_c$}

\author{E.-M. Ilgenfritz}
\affiliation{Joint Institute for Nuclear Research, VBLHEP,
            141980 Dubna, Russia}
\author{B.~V.~Martemyanov}
\affiliation{ Institute of Theoretical and Experimental Physics, 
            117259 Moscow, Russia}
\author{M.~M\"uller-Preussker}
\affiliation{Humboldt-Universit\"at zu Berlin, Institut f\"ur Physik,
            12489 Berlin, Germany}

\date{\today}

\begin{abstract}
Using a standard cooling method for $SU(3)$ lattice gauge fields  
constant Abelian magnetic field configurations are extracted after 
dyon-antidyon constituents forming metastable $Q=0$ configurations
have annihilated. These so-called Dirac sheets, standard and 
non-standard ones, corresponding to the two $U(1)$ subgroups of the 
$SU(3)$ group, have been found to be stable if emerging from the 
confined phase, close to the deconfinement phase 
transition, with sufficiently nontrivial Polyakov loop values.
On a finite lattice we find a nice agreement of the numerical 
observations with the analytic predictions concerning the 
stability of Dirac sheets depending on the value of the 
Polyakov loop.
\end{abstract}



\maketitle

\section{Introduction}
\label{sec:introduction}

In lattice gauge theories the cooling method is used to remove short distance
fluctuations in order to search for (approximate) classical solutions of the 
Euclidean field equations~\cite{Berg:1981nw,Ilgenfritz:1985dz,Hoek:1986nd,GarciaPerez:1993ki}. 
We consider this technique as a device~\cite{Ilgenfritz:2002qs}
(like smearing or filtering based on low-lying modes of the Dirac operator)
that may help to identify topological excitations generically present in the 
sample configurations representing the zero-temperature (or thermal) 
ensemble of gauge fields
~\cite{Ilgenfritz:2006ju,Bornyakov:2007fm,Bornyakov:2008im}.

Cooling studies of nonzero-temperature $SU(2)$ lattice fields 
~\cite{Ilgenfritz:2002qs} have identified as topological excitations both
calorons with nontrivial holonomy~\cite{Kraan:1998sn,Kraan:1998pm,Lee:1998bb} 
or dyon-antidyon pairs which finally annihilate. Sometimes this annihilation 
process provides a constant Abelian magnetic field called {\it Dirac sheet} 
(DS), which turns out to be either stable or unstable under further 
cooling~\cite{Ilgenfritz:2003fd}. The stability is strongly correlated with the 
spatial average value of the Polyakov loop (the holonomy) in the given stage 
of cooling. In \cite{Ilgenfritz:2004vx} an explanation for this observation 
was presented.

Some time ago we started cooling studies of $SU(3)$ gluodynamics,
applying the Cabibbo-Marinari procedure in the cooling mode for the 
standard Wilson action~\cite{Ilgenfritz:2005um}.  On plateaus characterized
by values of the action within the range $0.5 - 1.5$ times the one-instanton 
action $S_{inst}$ the emerging topological objects turned out to be either
calorons or anticalorons (dissociated or not dissociated into their three 
respective dyon or antidyon constituents) or one or two dyon-antidyon pairs.
Sometimes (similar to the $SU(2)$ case) the annihilation process of a 
dyon-antidyon pair leaves behind a constant Abelian magnetic field. 
In the $SU(3)$ case the structure of such Dirac sheets is somewhat richer 
than in the $SU(2)$ case. Below we will describe their analytic construction 
following a seminal paper by Gerard 't Hooft~\cite{'tHooft:1981sz}. We will 
expand the concept of marginal stability 
~\cite{GarciaPerez:1994kz,vanBaal:1995xk,vanBaal:1984ar} 
to the $SU(3)$ case. We shall find agreement between the analytically 
worked-out preconditions -- in terms of the holonomy -- 
for stability of the Dirac sheets in a finite volume on one hand
and the numerical observations for Monte Carlo generated -- and subsequently 
cooled -- lattice gauge fields.

\section{Dirac sheet solutions}
\label{sec:DS}

In lattice gauge theories usually periodic boundary conditions are applied
for the gauge fields (by default, if no special needs suggest something else).
Thus, the DS configurations, that can be obtained by the cooling procedure,
are periodic as well. 
The simplest way, however, to present analytic solutions with a constant 
color-magnetic field on a hypertorus uses twisted boundary 
conditions~\cite{'tHooft:1981sz}.
In this case most of the structure of the solutions is absorbed into twists
(the gauge transformations that the gauge fields acquire over the periods on 
a hypertorus). 
They look rather complicated and are even non-Abelian while the gauge fields 
themselves are rather simple. To have periodic solutions we should make clear 
that the twists can be removed by appropriate gauge transformations. 
The necessary condition for this is the commutativity of twists in different 
directions. Below we will apply this condition to find those solutions that 
allow to be made periodic.

Discussing the special selfdual solutions,  
't Hooft was considering the general $SU(N)$ case. The gauge field $A_\mu(x)$ 
and the field strength $F_{\mu\nu}(x)$ are strictly Abelian while the twists 
are non-Abelian. The gauge field $A_\mu(x)$ is proportional to
the diagonal traceless matrix $\omega = 2\pi~\mathrm{diag} (l,...,l,-k,...,-k)$
with positive integers $l$ and $k$ such that $l + k =N$
\bea
&A_{\mu}(x) = \omega \sum_{\nu} \alpha_{\mu\nu} x_{\nu}/L_{\mu}L_{\nu}\,, \nonumber \\
&F_{\mu\nu}(x) = -\omega (\alpha_{\mu\nu} -\alpha_{\nu\mu})/L_{\mu}L_{\nu}\,,
\eea
where $L_{\mu},~\mu=1,\ldots,4$ are the linear extensions of the hypertorus,
\beq
 \alpha_{\mu\nu} -\alpha_{\nu\mu} = n^{(2)}_{\mu\nu}/Nl-n^{(1)}_{\mu\nu}/Nk\,.
\eeq
The integers $n^{(2)}_{\mu\nu}$ and $n^{(1)}_{\mu\nu}$  
summed to $n_{\mu\nu} = n^{(1)}_{\mu\nu} +n^{(2)}_{\mu\nu}$
define the so-called twist tensor $n_{\mu\nu}$. 
For $n_{\mu\nu} = 0$ (mod. $N$) the twists are commuting and can be removed
by appropriate gauge transformations such that gauge fields become periodic.

For $n^{(2)}_{12}=-n^{(1)}_{12}=1$ 
(with other components equal to zero) $n_{\mu\nu} =0$, 
$\alpha_{12} -\alpha_{21} = 1/kl$ we get a constant magnetic field in the
third direction $B_3 = F_{12}$. The action of this field on the hypertorus
with $L_1=L_2=L_3=L_s$ and $L_4=L_t$ is equal to
\bea
S_{DS} &=& 1/2g^2(B^a_3)^2 V_4= 1/g^2Tr(B_3)^2 V_4\nonumber \\
&=& 8\pi^2/g^2 \times N/2kl \times L_t/L_s\,.
\eea 
Thus, for $SU(2)$ $S_{DS} = S_{inst}  L_t/L_s$, for $SU(3)$ 
$S_{DS} = 3/4 S_{inst}  L_t/L_s$,
where the instanton action is $S_{inst} = 8\pi^2/g^2 $.
In the $SU(2)$ case the magnetic field $B_3 $ is equal to 
$B_3 = 2\pi~\mathrm{diag} (1,-1)/L_s^2$ and its flux  $\Phi$ over the
$12$-plane of the hypertorus is a multiple of $2\pi$: 
$\Phi=2\pi~\mathrm{diag} (1,-1)$ . This means that in the periodic gauge 
such a field could remain Abelian because of  $\exp(i\Phi) = {\bf 1}\,.$
In the $SU(3)$ case the magnetic field $B_3 = \pi~\mathrm{diag} (1,1,-2)/L_s^2$ 
has a flux over the $12$-plane of the hypertorus equal to 
$\Phi=\pi~\mathrm{diag} (1,1,-2)$. Now $\exp(i\Phi) = \mathrm{diag} (-1,-1,1)$
is not equal to the unity matrix and this means that in the periodic gauge 
such a field could not remain Abelian. 

\section{$SU(2)$ embedded Dirac sheet solutions}
\label{sec:EDS}

The Dirac sheet seen on the lattice in the 
$SU(2)$ case ~\cite{Ilgenfritz:2003fd,Ilgenfritz:2004vx} is observed 
also in $SU(3)$ lattice simulations. We will call it {\it standard DS}.
New, specific for the $SU(3)$ case, is the Dirac sheet with an action value 
equal to $3/4$ of the action of the standard DS. In the following we will call 
it {\it non-standard DS}. It is also seen in lattice simulations.

In  $SU(2)$ a constant Abelian magnetic field is not stable under 
fluctuations of the gauge field. Charged (off diagonal) components of 
the gauge field have a Savvidy eigenmode~\cite{Savvidy:1977as} with negative 
eigenvalue
\beq
\lambda = -4\pi/L_s^2~.
\eeq
The situation can be stabilized by introducing a constant Abelian scalar 
potential $A^3_4$. Normally a constant Abelian scalar potential can be gauged 
away. In our case due to periodicity in time direction it can be gauged away 
only modulo $2\pi/L_t$.  The interaction of charged (off-diagonal) components 
of the gauge field with this potential adds a positive term to the
eigenvalue $\lambda$, turning it into
\beq
\lambda = -4\pi/L_s^2 +(A^3_4)^2~.
\eeq

The presence of the scalar potential leads to a nontrivial holonomy $H$ that 
is defined as
\beq
H=\lim_{|\vec x|\rightarrow\infty}
P\,\exp(i\int_0^{L_t} A_4(\vec x,t)dt).
\eeq
The holonomy is parametrized as 
$H = \mathrm{diag}(\mathrm{e}^{2\pi i\mu_1},\mathrm{e}^{2\pi i\mu_2})$
with $\mu_1 \leq \mu_2 \leq \mu_3 = 1+\mu_1$ and  $\mu_1+\mu_2 = 0$. Thus, 
positive numbers $m_1 = \mu_2 - \mu_1$, $m_2 = \mu_3 - \mu_2$ sum up to 
unity  $ m_1 + m_2 = 1$. The eigenvalue $\lambda$ then becomes equal to
\beq
\lambda = -4\pi/L_s^2 +(2\pi m_1/L_t)^2~,
\eeq
and its positiveness requires  $L_t/L_s\sqrt{\pi}< m_{1,2} < 1 - L_t/L_s\sqrt{\pi}$.
Therefore, nontrivial holonomy stabilizes DS and just this situation was 
observed in $SU(2)$ lattice cooling~\cite{Ilgenfritz:2003fd} and elucidated
in Ref.~\cite{Ilgenfritz:2004vx}.

Now let us consider the embedding of this standard DS event into $SU(3)$ group.
Let vector potentials $A_{1,2}$ be proportional to $\mathrm{diag}(1,-1,0)$ and 
the scalar potential to give the holonomy 
\beq
H = \mathrm{diag}(\mathrm{e}^{2\pi i\mu_1},\mathrm{e}^{2\pi i\mu_2}, 
               \mathrm{e}^{2\pi i\mu_3})
\eeq
with $\mu_1 \leq \mu_2 \leq \mu_3 \leq \mu_4 = 1+\mu_1$ and  
$\mu_1+\mu_2 +\mu_3 = 0$. Now three positive numbers 
$m_1 = \mu_2 - \mu_1$, $m_2 = \mu_3 - \mu_2$ , $m_3 = \mu_4 - \mu_3$ sum 
to unity  $ m_1 + m_2 + m_3  = 1$. Stability of the DS under fluctuations of 
charged (off-diagonal) $(1,2)-(2,1)$ components of the gauge fields requires  
$L_t/L_s\sqrt{\pi}< m_1 < 1 - L_t/L_s\sqrt{\pi}$. The other off-diagonal  
$(2,3)-(3,2)$ and  $(3,1)-(1,3)$ components of the gauge fields have charges
with respect to the $\mathrm{diag}(1,-1,0)$ generator of the $SU(3)$ group 
being two times smaller than the $(1,2)-(2,1)$ components.
Hence the stability of DS under their fluctuations requires
$L_t/L_s\sqrt{2\pi}< m_2 < 1 - L_t/L_s\sqrt{2\pi}$ and  
$L_t/L_s\sqrt{2\pi}< m_3 < 1 - L_t/L_s\sqrt{2\pi}$, correspondingly. 
Taking into account that the magnetic Abelian field  
could lie also in other $SU(2)$ subgroups of the $SU(3)$ group, i.e. 
would then be proportional to 
$\mathrm{diag}(1,0,-1)$ or to  $\mathrm{diag}(0,1,-1)$ generators,
we see that the standard DS in $SU(3)$ group will be stable 
for values of the holonomy restricted by the following constraints on the 
holonomy parameters $ m_1, m_2, m_3$ 
\beq
L_t/L_s\sqrt{2\pi}< m_{1,2,3} < 1 - L_t/L_s\sqrt{2\pi}\,.
\eeq

We shall visualize the stability criteria in a $(X,Y)$ plot in the complex
plane, $X = \Re(1/3~Tr~H)$ and $Y= \Im(1/3~Tr~H)$).
The corresponding region for the standard DS configurations is shown 
 on Fig. 1. The external curved triangle encloses all possible values 
of one third of the trace of an unitary matrix (the holonomy) that 
can be obtained by the variation of the phase parameters $ m_1, m_2, m_3$ 
in the region $0 < m_{1,2,3} < 1$, while the sum is constrained by
$ m_1 + m_2 + m_3  = 1$. 
The smaller, inscribed curved triangle (bounded by the dashed line) is the 
region of stability of standard DS events.
\begin{figure}
\includegraphics[width=.5\textwidth]{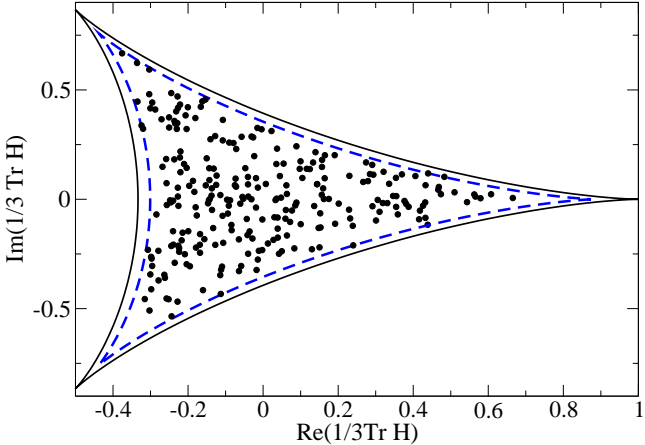}
\caption{The $SU(3)$ triangle and the inscribed region of stability 
expected for standard DS configurations (enclosed by the dashed line) 
compared with standard DS events found in actual lattice cooling 
(filled circles).}
\label{fig:standardds}
\end{figure}

\section{Non-standard Dirac sheets}
\label{sec:instability}

Coming now to the discussion of the stability of non-standard DS solutions
one should first stress that 
by construction constant Abelian magnetic fields can be supplemented only 
by a constant Abelian scalar potential proportional to the same diagonal  
$SU(3)$ generator to which the magnetic field is proportional.
If the magnetic field is equal to $B_3 = \pi~\mathrm{diag} (1,1,-2)/L_s^2$, 
then in a constant Abelian scalar potential 
\beq 
A_4 = \mathrm{diag} (2\pi\mu_1/L_t, 2\pi\mu_2/L_t, 2\pi\mu_3/L_t)
\eeq
the holonomy parameters $\mu_1$ and $\mu_2$ should be equal to each other:
$\mu_1 = \mu_2$ ($m_1 =0$ ).
The fluctuations of the $(1,2)-(2,1)$ components of gauge fields in this case 
do not interact with both the magnetic field and the static scalar potential. For  
fluctuations of charged  $(2,3)-(3,2)$ and  $(3,1)-(1,3)$ components 
the lowest modes have eigenvalues 
\beq
\lambda_{23} = -3\pi/L_s^2 +(2\pi m_2/L_t)^2~
\eeq
and 
\beq
\lambda_{13} = -3\pi/L_s^2 +(2\pi m_3/L_t)^2~
 \eeq
correspondingly. So, the stability of such non-standard DS solutions is possible for
\beq
m_1 = 0,~~~\sqrt{3/4\pi} L_t/L_s< m_{2,3} < 1 - \sqrt{3/4\pi} L_t/L_s~.
\eeq

For other non-standard DS solutions the region of stability can be obtained 
by the permutations of holonomy parameters $ m_1, m_2, m_3$. The stability 
region is shown in the $(X,Y)$ plot of Fig. 2 and happens to coincide with 
the boundary of the unclosed $SU(3)$ triangle of Fig. 1.
\begin{figure}
\includegraphics[width=.5\textwidth]{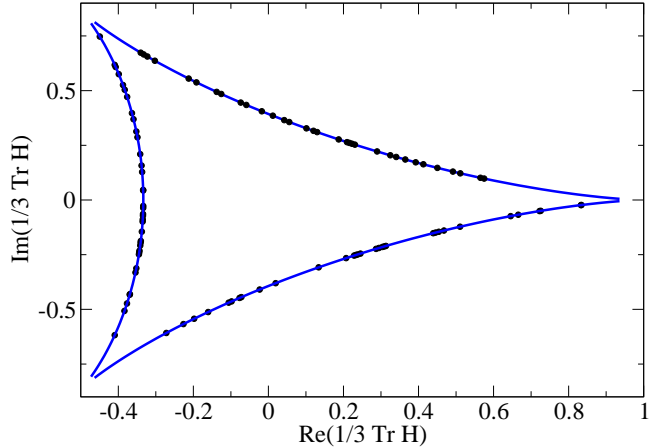}
\caption{The region of stability of non-standard DS configurations (the three 
sides of the unclosed $SU(3)$ triangle)
compared with non-standard DS events found in actual lattice cooling 
(filled circles).}
\label{fig:nonstandardds}
\end{figure}

\section{Numerical results}
\label{sec:numericalresults}

For a numerical study of standard and non-standard DS solutions 
we have employed the standard Wilson plaquette action $S_W$, creating
an ensemble with $\beta = 6/g^2 \,$ where $~g$ denotes the bare coupling 
constant. On a lattice for $L_t=4,L_s=16$  the coupling constant related 
to the first order deconfinement transition
is equal to $\beta_d \simeq 5.69$. 
The initial Monte Carlo ensemble 
was generated in the confined phase at $\beta = 5.63$. 
As expected, this has guaranteed that in the process of cooling the 
holonomy has remained sufficiently non-trivial, such that the emerging
DS configurations were stable.
We have found configurations stable against 
further cooling with the action $S=1/4 S_{inst}$ and $S=3/16 S_{inst}$ in
perfect agreement with analytical knowledge.
We have stopped cooling at the moment, when the relative
variation of action density inside the configuration became smaller than 
$10^{-4}$ (homogeneous configurations) and have measured
the value of holonomy (the average Polyakov loop). 
The Polyakov loop also has happened homogeneous. The distance
of local values of it from the average value was not larger than $10^{-5}$.
The scatter plots of DS events in the $(X,Y)$ plane of the real and
imaginary part of the Polyakov loop are shown in Figs. 1 and 2. 
The dots lie perfectly inside the regions 
of stability for the respective type of DS configurations.
The configurations obtained turned out to be purely magnetic and -
applying maximally Abelian gauge -- show constant Abelian magnetic fluxes.

We did not particularly attempt to find Dirac sheets at higher temperature,
$\beta > \beta_c$. We know from other simulations that the holonomy of
such equilibrium configurations under cooling rapidly evolves towards central 
elements where Dirac sheets are unstable and therefore would have escaped 
observation.

\section{Conclusion}
\label{sec:conclusion}
In conclusion, purely Abelian constant magnetic field configurations 
have been observed emerging from the process of cooling equilibrium 
(Monte Carlo) lattice fields representing the confined phase of $SU(3)$ 
gluodynamics. They were found to be absolutely stable provided their 
Polyakov loop was sufficiently non-trivial. We have shown here that this 
fact is related to the notion of marginal stability of the appropriate 
constant magnetic field configurations.

Finally we have to admit that the Dirac sheet configurations discussed
in this paper will not play any r\^ole in the thermodynamic limit of the
theory since their action tends to zero in this limit. 

\subsection*{Acknowledgments}
We thank our collaborator V.K. Mitrjushkin for drawing our attention to 
the extremely stable plateaus occuring during very long cooling trajectories 
in the $SU(3)$ case.   
B.V.M. gratefully acknowledges the kind hospitality extended to him at 
the Physics Department of Humboldt-University Berlin.


\end{document}